\documentclass[11pt,a4paper]{article}
\pdfoutput=1

\usepackage[left=0.8in,right=0.8in,top=1.0in,bottom=1.2in]{geometry}

\usepackage{mathrsfs,graphicx,rotating,amsmath,amsfonts,mathtools,booktabs,amssymb,wasysym,caption}
\usepackage{cite}
\usepackage{hyperref}
\usepackage{authblk}
\usepackage{slashed}
\usepackage[table,xcdraw,dvipsnames]{xcolor}
\usepackage{graphicx}
\usepackage{bbold}
\usepackage[utf8x]{inputenc}
\usepackage[english]{babel}
\usepackage{multirow,multicol}
\usepackage{epstopdf}
\usepackage{bbm}
\usepackage{changepage}
\usepackage{appendix}
\usepackage{libertine}
\usepackage{braket}
\usepackage{wasysym}
\usepackage{empheq}
\usepackage{cancel}
\usepackage{enumitem}
\usepackage{mathrsfs}
\usepackage{lmodern}
\usepackage{tabularx}
\usepackage{multicol}
\usepackage{multirow}
\usepackage{color}
\usepackage{mathtools}
\usepackage{verbatim}
\usepackage{arydshln}
\usepackage{url}
\usepackage[normalem]{ulem}
\usepackage{youngtab}

\newcommand{\cB}{{\mathcal B}}

\newcommand{\gsim}{\lower.7ex\hbox{$\;\stackrel{\textstyle>}{\sim}\;$}}
\newcommand{\lsim}{\lower.7ex\hbox{$\;\stackrel{\textstyle<}{\sim}\;$}}

\title{\textbf{Minimal flavour deconstruction}}
 
\author[1]{Riccardo Barbieri\thanks{riccardo.barbieri@sns.it}}
\author[2]{Gino Isidori\thanks{isidori@physik.uzh.ch}}
\affil[1]{Scuola Normale Superiore, Piazza dei Cavalieri 7, 56126 Pisa, Italy}
\affil[2]{Physik-Institut, Universit\"at Z\"urich, CH-8057 Z\"urich, Switzerland}

\begin{document} 
\maketitle

\abstract{ 
We construct two concrete examples of flavour non-universal gauge theories which, after the inclusion of all $d\leq 4$ gauge invariant operators, allow to describe the observed pattern of flavour in the charged fermion sector without any small Yukawa coupling ($y \gtrsim 0.1$). Guided by the criterium of minimality, we assume that flavour non universality is confined to the Abelian sector of the gauge group: the universal hypercharge emerges after a sequence of symmetry-breaking steps characterised by two high mass scales,  $\Lambda_{[23]} < \Lambda_{[12]}$, where the  second and the first fermion generations  get their mass respectively. At least in one of  the two models the smaller of these scales can be in the 10 TeV range, consistently with current bounds from flavour observables. Both models are extended to include as well neutrino masses and mixings.
}

\tableofcontents 

\section{Motivations and purpose}
\label{Intro}

The {\it flavour puzzle}, together with the {\it hierarchy problem}, conceivably related to each other via the Yukawa couplings, represent the most serious ``structural" problems of the otherwise extraordinarily successful Standard Model (SM) in its comparison with all particle physics experiments so far. Ideally, to solve the flavour puzzle would mean to make calculable the thirteen vastly spread free parameters in the charged fermion flavour sector, even better if extended to include the neutrino parameters, in terms of definitely fewer inputs. In the hope to pave the way to such a major step, here we attack a  less pretentious problem:  building an explicit gauge theory which, after the inclusion of all $d\leq 4$ gauge invariant operators, allows to describe the observed pattern of flavour  without any small Yukawa coupling  ($y \gtrsim 0.1$). The way this is obtained is by assuming that flavour universality of gauge interactions is only a low energy property: the underlying gauge structure is manifestly flavour non-universal. This {\em flavour-deconstructed} gauge symmetry is broken to the universal subgroup in two steps, characterised by two mass scales,  $\Lambda_{[23]} < \Lambda_{[12]}$ (both higher that the electroweak scale), where the  second and the first fermion generations  get their mass respectively. 

The hypotheses of flavour non-universal interactions and that light-generation masses are generated at 
scales well above the electroweak one, is not new and has been pursued in different contexts (see e.g.~\cite{Li:1981nk,Barbieri:1983uq,Dvali:2000ha,Craig:2011yk,Panico:2016ull}).
In a sense technicolour can be seen as a progenitor of this  idea~\cite{Farhi:1980xs}.  An additional motivation which has arisen relatively more recently is related to the emergence of the {\it little hierarchy problem}: the lack of new physics (NP) signals from direct searches are particularly stringent for new degrees of freedom coupled strongly to the light families (both quarks and leptons), whereas they do not exceed $1\div2$~TeV for new dynamics involving mainly the third generation
(see e.g.~\cite{Allwicher:2023shc}). Hence new dynamics  involving mainly the third generation is the best option for addressing both the flavour puzzle and the hierarchy problem~\cite{Barbieri:2021wrc,Davighi:2023iks}.  Beside these general considerations, a renewed phenomenological interest in flavour non-universal gauge interactions has also been triggered by a series of hints of deviations from SM in $B$ decays (see e.g.~\cite{Koppenburg:2023ndc}).  While none of the hints is particularly compelling nowadays, the  model-building efforts presented in~\cite{Bordone:2017bld,Greljo:2018tuh,Fuentes-Martin:2020pww,Fuentes-Martin:2020bnh,Davighi:2022fer,Fuentes-Martin:2022xnb,FernandezNavarro:2022gst,Davighi:2022bqf}
are interesting examples of the possibility to accomodate low-scale NP  via non-universal gauge interactions connected to the origin of the fermion mass hierarchies.

Leaving aside the hints of deviations from the SM, which are currently subject to further experimental verification, 
the scope of this work is to describe the observed spectrum of quarks and leptons via a minimal flavour non-universal extension of the SM gauge group.
In particular, we insist on having a fully renormalizable theory in four dimensions without small Yukawa couplings. We identify a unique choice of 
non-universal $U(1)$ gauge groups, and a related  symmetry breaking pattern, fulfilling this goal. Less obvious is the choice for the 
field content, in particular the vector-like fermions, which mediate the symmetry breaking to the chiral fermions: we investigate 
two alternative options in this respect.
Analysing these two concrete options allows us to distinguish general features from model-dependent conclusions when deriving 
bounds on the different energy scales from present data. We also compare our results to the recent proposals in~\cite{FernandezNavarro:2023rhv,Davighi:2023evx},  where different versions of flavour-deconstructed hypercharge
have been discussed. 

Flavour-changing  observables turn out to be an essential probe of this class of class of models. 
At least in one of the two options present data  allow the lowest new states
(associated to the $\Lambda_{[23]}$ scale)  to be in the  few TeV range. In both options,
the potential of the precision program in flavour physics of the next $10\div 15$ years,  expected to improve in precision by about 
one order of magnitude 
in several observables,  emerges in all evidence.

The paper is organised as follows: in Section~\ref{TCM} we introduce and motivate the non-universal gauge group and the two options for the additional (non-standard) matter content.
In Section~\ref{BSNP} we first analyse the structure of the Flavour Changing Neutral Currents (FCNCs) in both models, and analyse the corresponding bounds 
from current flavour observables. In Section~\ref{ETN} we propose two extensions of the matter content able to describe also the observed neutrino masses and mixings. 
The results are summarised in the Conclusions. In the Appendix we discuss problems with alternative choices for the non-universal $U(1)$ gauge groups.

\section{Two concrete models }
\label{TCM}

\begin{figure}[t]
\centering
\includegraphics[clip,width=.95\textwidth]{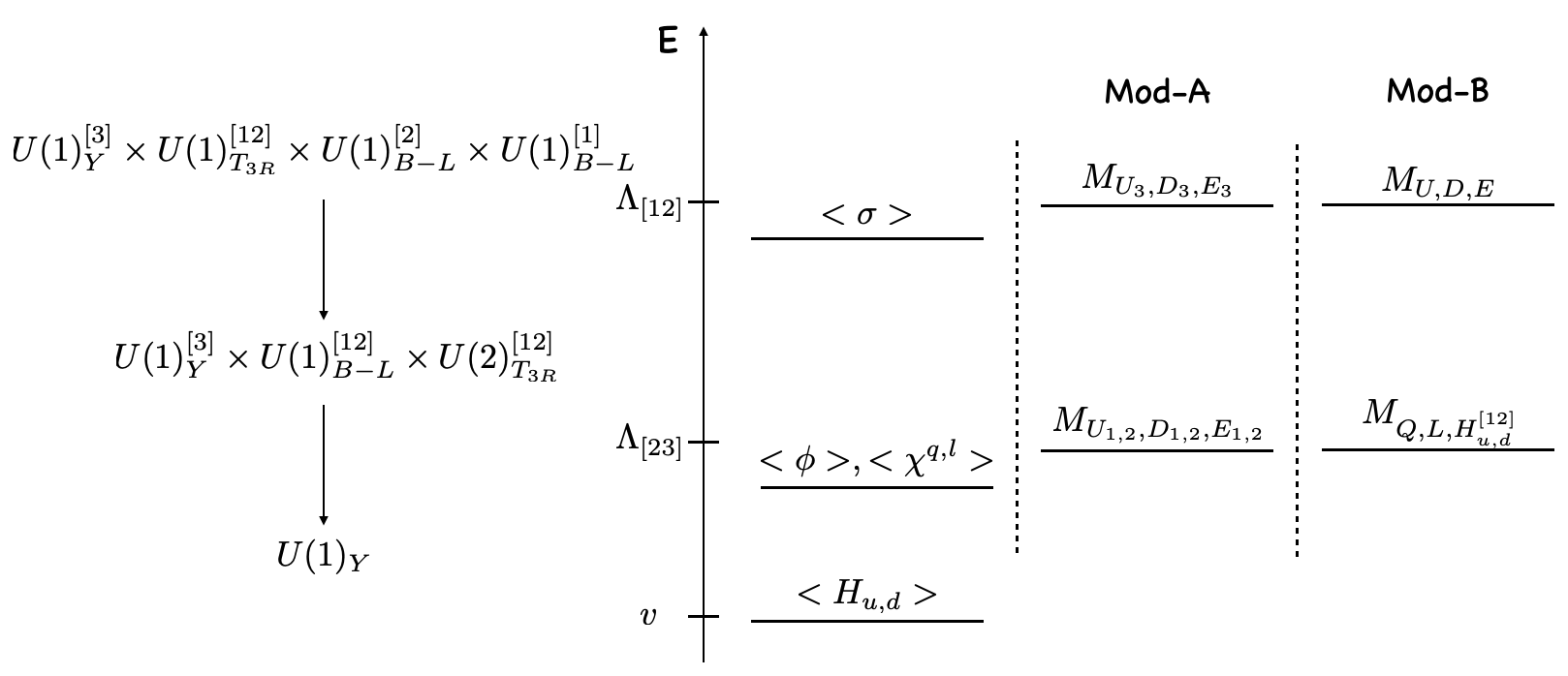}
\caption{\small  Overall representation of Model A and Model B}
\label{fig:Fig4overall}
\end{figure}

The concrete examples we consider are based on the gauge group
\begin{equation}
G = SU(3)\times SU(2)\times U(1)_Y^{[3]} \times U(1)_{B-L}^{[12]} \times U(1)_{T_{3R}}^{[2]} \times U(1)_{T_{3R}}^{[1]}
\label{eq:G}
\end{equation}
where $SU(3)$ and $SU(2)$ act universally on the three fermion families, as in the SM, whereas the  $U(1)$ groups 
act non-universally only on one or two families, as indicated by the corresponding superscripts.
The way the gauge group acts on the three sixteen-plets of chiral fermions, including the right handed neutrinos needed to cancel gauge anomalies, is self-explanatory. 
The symmetry breaking cascade of the $U(1)$ factors to the SM $U(1)_Y$ , which is common to the two examples, 
\begin{equation}
U(1)_Y^{[3]} \times U(1)_{B-L}^{[12]} \times U(1)_{T_{3R}}^{[2]} \times U(1)_{T_{3R}}^{[1]} 
\stackrel{\langle\sigma\rangle}{\longrightarrow}
U(1)_Y^{[3]} \times U(1)_{B-L}^{[12]} \times U(1)_{T_{3R}}^{[12]} 
\stackrel{\langle\phi,\chi\rangle}{\longrightarrow}
 U(1)_Y
\end{equation}
is achieved via the set of scalar fields reported in Table~\ref{tab:SSB} (see also Figure~\ref{fig:Fig4overall}).
Given the tight bounds about non-universality among light families, we assume a two scale structure with $\langle \sigma \rangle$ significantly larger than $\langle \phi \rangle 
\sim \langle \chi^{q,l} \rangle$ \footnote{Note that the scalars $\chi^q$ and $\chi^l$, which will control the 
light-heavy left-handed mixing among quarks and leptons respectively, admit a $d=4$ gauge invariant term in the potential of the type $\chi^l(\chi^{q})^3$. Hence no pseudo-Goldstone boson is generated in the breaking of the gauge group. }. The Higgs doublets $H_u$ and $H_d$, which are responsible for the breaking of the electroweak symmetry, 
are distinguished by a softly broken $Z_2$ symmetry which makes them couple to the up-type quarks/neutrinos  and to the down-type quarks/charged leptons respectively. We assume $\tan\beta = v_u/v_d = 10\div 10^{2}$ to justify the
overall normalisation of  the different Yukawa couplings. We also assume negligible kinetic mixings between the different $U(1)$ gauge bosons, as they could arise from a suitable semi-simple gauge group at higher energies.

\begin{table}[t]
 $$\begin{array}{c|c||c|c|c|c|c}
{\rm vev\ scale}  & {\rm Field} &  U(1)_Y^{[3]} & U(1)_{B-L}^{[12]} & U(2)_{T_{3R}}^{[2]}& U(1)_{T_{3R}}^{[1]}& SU(3)\times SU(2)\\ \hline
 v & H_{u,d} & -1/2& 0 & 0 & 0 &(\bold{1},\bold{2}) \\ \hline \hline
\multirow{3}{*}{ $ O(10^{-1}) \times \Lambda_{[23]}$ }  & \chi^q   &-1/6& 1/3 & 0 &0 & (\bold{1},\bold{1}) \\ \cline{2-7}
& \chi^l & 1/2& -1 & 0 &0 & (\bold{1},\bold{1}) \\ \cline{2-7}
& \phi & 1/2& 0 & -1/2& 0&(\bold{1},\bold{1}) \\ \hline \hline
O(10^{-1})  \times \Lambda_{[12]} & \sigma &0 & 0 & 1/2 & -1/2  & (\bold{1},\bold{1}) \\ \hline 
\end{array}$$
\caption{\small  Scalar fields responsible for the symmetry-breaking pattern of the gauge group.}
\label{tab:SSB}
\end{table}

The choice of the gauge group (\ref{eq:G}) and of the symmetry-breaking pattern can be understood by looking at the structure of the Yukawa couplings
\begin{equation}
\mathcal{L}_Y = H_u \bar{q}Y_u u + H_d^* \bar{q}Y_d d + H_d^* \bar{l}Y_e e.
\end{equation}
On general grounds, we aim at achieving the following structure
\begin{equation}
Y \sim \left(\begin{array}{ccc}
	y_1 	&  y_2 \lambda &  y_3 \lambda^3 	\\
	- 	&  y_2 		&  y_3 \lambda^2 	\\
	-	&  - 			&  y_3 
\end{array}\right)
\label{eq:Ygen}
\end{equation}
for each Yukawa coupling, where $\lambda = |V_{us}| \approx 0.2$, $y_1\ll y_2 \ll  y_3 $ and the entries denoted by `$-$', responsible for right-handed mixing,  not observable 
in the SM,\footnote{Right-handed mixing becomes observable in extensions of the SM such as those 
we are considering here. In this context, given the strong experimental constraints on flavour violations involving 
right-handed fermions, a safe phenomenological choice is obtained requiring all the `$-$' entries in (\ref{eq:Ygen}) to be at most of $O(y_2\lambda^2)$.}  are not larger than the corresponding off-diagonal terms explicitly shown.
In the quark case, the size of the explicit off-diagonal terms is inferred from the structure of the 
Cabibbo-Kobayashi-Maskawa (CKM) matrix, which  
provides an information about the (left-handed) misalignment between up- and down-quark Yukawa couplings.

As discussed in~\cite{Davighi:2023iks}, once we assume a universal $SU(2)$ group, the request of allowing only 
third-generation Yukawa couplings via the gauge symmetry is obtained via the  flavour-deconstruction of 
$U(1)_{B-L}$ (acting on both left-handed and right-handed fermions) and $U(1)_{T_{3R}}$
(acting on right-handed fermions only).
If we also aim at explaining the hierarchy $y_1\ll y_2 $,  we are forced to 
the further deconstruction of $U(1)_{T_{3R}}$ among the first two families. 
On the other hand, given that $\lambda$ is not a
very small parameter, there is no need to distinguish the  first two families of left-handed quarks.
This is the rationale behind the choice of the gauge group in Eq.~(\ref{eq:G}).

Having chosen the gauge group and the symmetry-breaking scalars in Table~\ref{tab:SSB},
the parametric structure of the Yukawa couplings deduced by a general spurion analysis is:
\begin{equation}
\begin{array}{c:c:c:cc}
	& \multicolumn{2}{c:}{U(1)_{B-L}^{[12]}}  		&     	&      \\[2pt] 
	& U(1)_{T_{3R}}^{[1]} & U(1)_{T_{3R}}^{[2]} 	&  	&      \\[5pt] \cdashline{2-5}   
   \multirow{3}{*}{  $Y \sim \left(\begin{array}{c}  ~   \\  ~  \\  ~  \end{array}\right. $\hskip -0.8cm  } \raisebox{8pt}{ \phantom{a} }	
   	& \multirow{2}{*}{$O(\epsilon_\sigma \epsilon_\phi)$ }	&  \multirow{2}{*}{$O(\epsilon_\phi)$ }	&   
   \multirow{3}{*}{  $\left. \begin{array}{c}  \raisebox{-7pt}{$O(\epsilon_\chi)$}  \\   ~  \\  O(1)  \end{array}\right)$}
	&   \multirow{2}{*}{   $U(1)_{B-L}^{[12]}$  }       \\[3pt]
 	&  		&  		&	    	&    	\\   \cdashline{2-5}
 	&   	 \raisebox{-2pt}{$O(\epsilon_\sigma \epsilon_\phi \epsilon_\chi)$} 	&     
	 	 \raisebox{-2pt}{$O(\epsilon_\phi \epsilon_\chi)$} 	&     		&      \\
\end{array}  \\[5pt]
\label{eq:Ypar}
\end{equation}
Here the $\epsilon_i$ denote the different suppression factors 
due to the vacuum expectation value (vev) of each scalar field (or of its complex conjugate)
\begin{equation}
\epsilon_\sigma = \frac{ \langle \sigma \rangle}{ \Lambda_{[12]} }\,, \qquad 
\epsilon_{\phi } =  \frac{ \langle  \phi  \rangle}{ \Lambda_{[23]} }\,, \qquad
\epsilon_{\chi} =  \frac{ \langle  \chi^{q,l} \rangle}{ \Lambda_{[23]} }\,,
\label{eq:epsilon}
\end{equation}
with $\Lambda_{[12]}  \gg \Lambda_{[23]}$.  The  two examples we consider 
differ in the heavy matter content, i.e.~in the explicit fields with masses of $O(\Lambda_{[23]})$ and $O(\Lambda_{[12]})$,
hence in the explicit calculation of the coefficients in Eq.~(\ref{eq:Ypar}), 
but not in the general parametric structure.

Before discussing in detail the two models, we comment on the differences between the choice in Eq.~(\ref{eq:G})
and other recent attempts to justify the hierarchical structure of the effective Yukawa couplings in terms of 
flavour non-universal gauge groups. In this work we do not require the embedding of the $U(1)$
groups into a semi-simple group, at least at low scales, while we are driven by the minimality of the gauge group 
yielding a phenomenologically viable Yukawa structure.  This is why, contrary to \cite{Davighi:2023iks},
we combine $U(1)_{B-L}^{[3]}$ and  $U(1)_{T_{3R}}^{[3]}$ into the more minimal $U(1)_Y^{[3]}$ (where $Y^{[3]} = (B-L)^{[3]}/2 + T_{3R}^{[3]}$).
A further reduction cannot be  applied to the groups acting on the light families if we aim 
achieving {\em both} $y_1 \ll y_2 $ and a sizeable mixing among the light families, at least without tuning of the couplings 
or introducing extra {\it ad hoc} scalars.  This is why we consider less interesting the more minimal choices  $U(1)_Y^{[3]} \times U(1)_Y^{[2]} \times U(1)_Y^{[1]}$
 and $U(1)_Y^{[3]} \times U(1)_Y^{[12]}$ which have been discussed in~\cite{FernandezNavarro:2023rhv} 
 and~\cite{Davighi:2023evx}, respectively. 
 To illustrate the problems encountered in these cases we outline in Appendix A an explicit model based on $U(1)_Y^{[3]} \times U(1)_Y^{[2]} \times U(1)_Y^{[1]}$.

\subsection{Model A}
\label{MA}

The fermion content of Model A includes, other than the standard chiral fermions, three generations of Vector-Like (VL) $SU(2)$-singlets 
as shown in Table \ref{Table:VLFA}.  
The full set of Yukawa-like couplings and mass terms is determined by the transformation properties of these fields. For example in the up-quark sector $(i=1,2, \alpha = 1,2)$
\begin{eqnarray}
\mathcal{L}^u_Y &=&  (y^u_3\, \bar{q}_{3} u_{3} H_u +  y^u_{i\alpha}\,  \bar{q}_{i} U_{\alpha}  H_u 
+  y^{\chi_u}_{\alpha}\,   \bar{U}_{\alpha} u_{3}  \chi^q + y^{\phi_u}_{\alpha 2}\,  \bar{U}_{\alpha} u_{2} \phi  
+ y^{\phi_u}_{\alpha 3}\,    \bar{U}_{R\alpha} U_{L3}  \phi \nonumber \\
&&  +\,  \hat{y}^{\phi_u}_{\alpha 3}\,    \bar{U}_{L\alpha} U_{R3}  \phi  + y^{\sigma_u}_1\,  \bar{U}_{3} u_{1} \sigma+ {\rm h.c.} ) + M_{U_3}\, \bar{U}_{3}U_3 + M_{U_\alpha} \,   \bar{U}_{\alpha}{U}_{\alpha}
\label{LYA}
\end{eqnarray}
where, unless specified, the chirality component is left understood since non ambiguous $(q \equiv q_L, u \equiv u_R)$.
The dimensionless couplings in front of each term are understood to be in the range ${0.1 \div 1}$.
In the down-quark  sector everything is similar, but for the replacement  $H_u \to H_d$, as in the charged-lepton sector where also $\chi^q \to \chi^l$.
The overall picture of Model A is represented in Fig.~\ref{fig:Fig4overall}

\begin{figure}[t]
\centering
\includegraphics[clip,width=.85\textwidth]{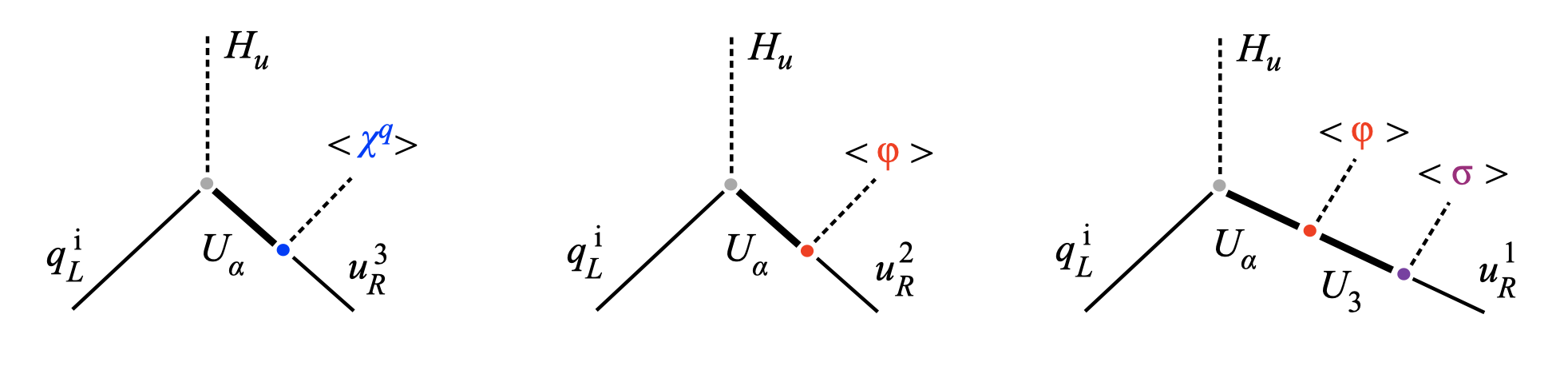}
\caption{\small  Diagrams describing sub-leading contributions to the up-quark Yukawa coupling in Model A}
\label{fig:ModAYukawa}
\end{figure}

To a sufficient level of approximation  the true Yukawa couplings of the massless fermions, before electroweak symmetry breaking, 
are obtained by integrating out first the heavy VL fermions $\{U_3, D_3, L_3\}$, and then 
$\{U_\alpha, D_\alpha, L_\alpha\}$, as indicated in Figure~\ref{fig:ModAYukawa} in the up-quark case.
Setting $M_{F_1}\approx M_{F_2} \equiv M_{[23]}  \ll  M_{F_3}  \equiv M_{[12]}$ ($F=U,D,E$)  one recovers the parametric structure in Eq.~(\ref{eq:Ypar}) with  $\Lambda_{[23]} = M_{[23]} $ and   $\Lambda_{[12]} = M_{[12]}$.
 
 \begin{table}[t]
 $$\begin{array}{cc|c|c|c|c|c}
& & U(1)_Y^{[3]} & U(1)_{B-L}^{[12]} & U(2)_{T_{3R}}^{[2]}& U(1)_{T_{3R}}^{[1]}& SU(3)\times SU(2)\\ \hline
\multirow{3}{*}{  $\begin{array}{c} {\rm light\  VL} \\  (\alpha=1,2) \end{array}$   }
& U_\alpha & 1/2& 1/3 & 0 &0 &(\bold{3},\bold{1}) \\ \cline{2-7}
& D_\alpha & -1/2& 1/3 & 0 & 0&(\bold{3},\bold{1}) \\ \cline{2-7}
& E_\alpha & -1/2& -1 & 0 & 0&(\bold{1},\bold{1}) \\ 
 \hline\hline
\multirow{3}{*}{ heavy\ VL   }  
&U_3 & 0& 1/3 & 1/2 &0 &(\bold{3},\bold{1}) \\ \cline{2-7}
&D_3 & 0& 1/3 & -1/2 &0 &(\bold{3},\bold{1}) \\ \cline{2-7}
&E_3 & 0& -1 & -1/2 & 0&(\bold{1},\bold{1}) \\ \hline
\end{array}$$
\caption{\small  Non-standard matter content of Model A. }
\label{Table:VLFA}
\end{table}

 In the limit of degenerate $M_{F_\alpha}$, due to the global $U(2)$ symmetries acting on $U_\alpha, D_\alpha, E_\alpha$ in the gauge sector,  we can always choose a flavour basis where  $y^{\chi_f}_{1}=0$ (for each $f=u,d,e$).
 In such basis, the  up-type Yukawa coupling assumes the form
 \begin{equation}
 Y_u  \approx  \begin{pmatrix}
  y^u_{1\alpha} \hat{y}^{\phi_u}_{\alpha 3}  y_1^{\sigma_u} \epsilon_\sigma \epsilon_{\phi }  &    y^u_{1\alpha} y^{\phi_u}_{\alpha 2} \epsilon_\phi  & y^{u}_{12}  y^{\chi_u}_{2} \epsilon_\chi   \\
   y^u_{2\alpha}  \hat{y}^{\phi_u}_{\alpha 3}  y_1^{\sigma_u} \epsilon_\sigma \epsilon_{\phi } & y^u_{2\alpha} y^{\phi_u}_{\alpha 2} \epsilon_\phi  &   y^{u}_{22}  y^{\chi_u}_{2}  \epsilon_\chi  \\
 \approx 0 &  \approx 0 & y^u_3
 \end{pmatrix}
  \label{eq:YuA}
 \end{equation}
 and similarly for $Y_{d,e}$. These matrices are approximately diagonalized by a unitary transformation in the left sector only, 
leading to the following eigenvalues
\begin{equation}
m_t\approx  y^u_3 v_u\,, \quad\  m_{b(\tau)} \approx  y^{d(e)}_3 v_d\,, \quad\  
\frac{m_{f_2}}{m_{f_3}}  \approx  \frac{y^f_{2\alpha} y^{\phi_f}_{\alpha 2} }{y^f_3} \epsilon_\phi\,,
\quad\
\frac{m_{f_1}}{m_{f_2}}\approx   \frac{  y^f_{1\alpha} y^{\phi_f}_{\alpha 3}}{ y^f_{2\alpha} y^{\phi_f}_{\alpha 2} } y^{\sigma_f}_1     \epsilon_\sigma\,.
\end{equation}
Concerning quark-flavour mixing,  the small $3\to2$ mixing takes the form  
\begin{equation}
|V_{cb}| \approx  \theta^d_{32} -  \theta^u_{32}  \approx  \left( y^{d}_{22} \frac{y^{\chi_d}_{2}}{y^d_3}  -  y^{u}_{22} \frac{y^{\chi_u}_{2} }{y^u_3} \right) 
 \epsilon_\chi\,,
\end{equation}
where we have assumed the mild hierarchy $y^{u}_{12}/y^{u}_{22} 
 =O(\lambda)$, and 
similarly in the down sector. Note that the degree of down-alignment of the third generation can be controlled by a single 
Lagrangian parameter:   both $\theta^d_{32}$  and $\theta^d_{31}$ vanish in the limit $y^{\chi_d}_2 \to 0$.

\subsection{Model B}
\label{MB}

The additional matter content of Model B is shown in Table \ref{tab:VLFB}.  Contrary to Model A, in this case also the Higgs sector is extended. Moreover, the vector-like fermions include both $SU(2)_L$ singlets and $SU(2)_L$ doublets.  
The complete Yukawa Lagrangian can be decomposed as
\begin{equation}
\mathcal{L}_Y = \mathcal{L}^u_Y + \mathcal{L}^d_Y + \mathcal{L}^e_Y  +\mathcal{L}^{Q,L}_Y
\end{equation}
where 
\begin{eqnarray}
\mathcal{L}^u_Y &=&  (y^u_3\, \bar{q}_{3} u_{3} H_u  +  y^{u}_{Q} \, \bar{Q} u_{3} H_u  + 
    y^{u}_{i2} \, {\bar q}_i u_{2} H^{[12]}_u 
+   y^{u}_{i1} \, {\bar q}_i U  H^{[12]}_u  
+  y^{\sigma_u}_1\,  \bar{U}u_{1} \sigma \nonumber \\
&& +   \lambda^\phi_u\,  M_H  H^{[12]}_u H^\dagger_u \phi\, +{\rm h.c.} )  + M_U\, \bar{U}U\,,  
\label{LYB}
\end{eqnarray}
\begin{equation}
\mathcal{L}^{Q,L}_Y =  (y^{\chi_q}_{ i }\,   \bar{q}_i  Q  \chi^q  + y^{\chi_l}_{ i }\,   \bar{\ell}_i  L  \chi^l\, +{\rm h.c.} )   + M_Q \,  \bar{Q} {Q}
+ M_L\,  \bar{L} {L}\,,
\end{equation}
and with $\mathcal{L}^{d,e}_Y$ fully analogous to $\mathcal{L}^{u}_Y$ with an appropriate change of fermion and 
scalar fields. 
In this model $q_3$ and the left-handed component of $Q$ have exactly the same 
gauge charges. This leads to a global $U(2)$ symmetry acting on $\{q_3, Q_L \}$ in the gauge sector:
we have used this symmetry to eliminate a possible vector-like mass term of the type $M_Q^\prime \bar q_3  Q$.
The same applies to the lepton doublets $\ell_3$ and $L_L$ and, analogously in the $SU(2)$-singlet sector, to 
$\{u_2, U_{R}\}, \{d_2, D_{R}\}, \{e_2, E_{R}\}$. The overall picture of Model B is represented in Fig.~\ref{fig:Fig4overall}

\begin{figure}[t]
\centering
\includegraphics[clip,width=.85\textwidth]{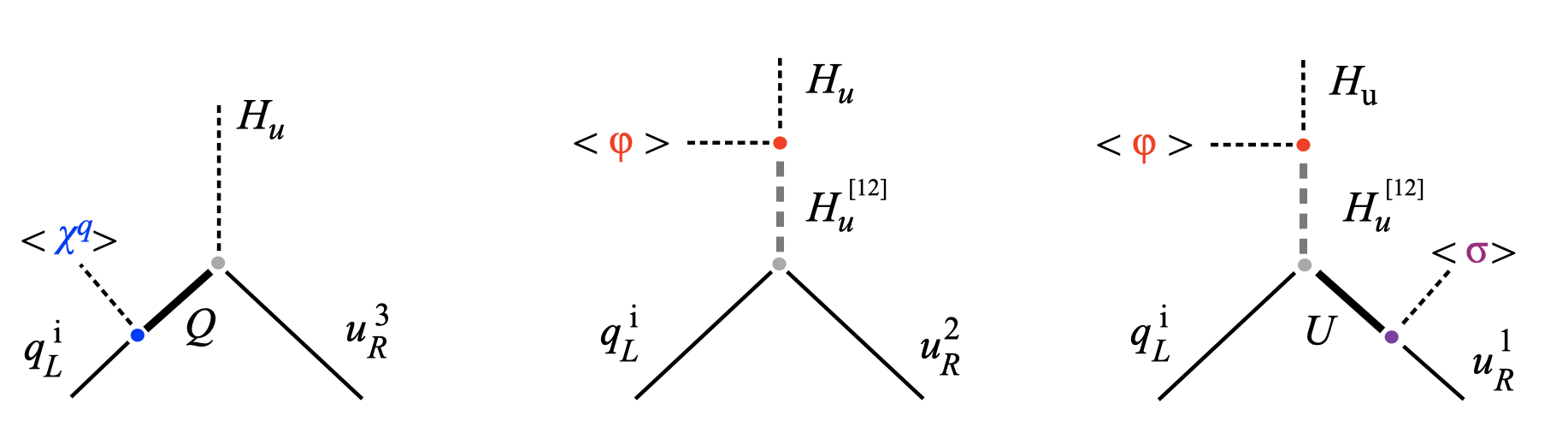}
\caption{\small  Diagrams describing sub-leading contributions to the up-quark Yukawa coupling in Model B}
\label{fig:ModBYukawa}
\end{figure}

The  Yukawa couplings for the massless fermions are obtained by integrating 
out first the $\{U, D, E\}$ fermions, as in Model A,
and then the $SU(2)_L$-charged VL fermions and the heavy Higgs fields  ($H^{[12]}_{u,d}$), 
as illustrated in Figure~\ref{fig:ModBYukawa}. 
For simplicity, we  assume a common mass $M_H = O(1) \times M_Q$  for the Higgs fields.
The parametric structure in Eq.~(\ref{eq:Ypar}) is then recovered for  $\Lambda_{[12]} = M_F\equiv M_{[12]} $ ($F=U,D,E$) and 
$\Lambda_{[23]} = M_{Q} \approx M_L\equiv M_{[23]}  $.

Since the Yukawa couplings in the two models have  same parametric dependence 
from the small ratios $\{\epsilon_\sigma, \epsilon_\phi, \epsilon_\chi\}$, 
the expressions for physical masses and mixing angles differ only because of different 
combinations of the $y_i$ couplings. Thanks to the global $U(2)$ symmetry acting on $\{q_1, q_2\}$, without loss of generality we can 
 set  $y^{\chi_q}_{1}=0$, such that $Y_u$ assumes the form 
  \begin{equation}
 Y_u  \approx  \begin{pmatrix}
  y^u_{11}  y_1^\sigma  \lambda^\phi_u \epsilon_\sigma \epsilon_{\phi }    &  y^{u}_{1 2}  \lambda^\phi_u  \epsilon_\phi  & 0   \\
   y^u_{21}  y_1^\sigma  \lambda^\phi_u \epsilon_\sigma \epsilon_{\phi } & y^u_{22} \lambda^\phi_u \epsilon_\phi    &   y^u_{Q}  y^{\chi_q}_{2}  \epsilon_\chi  \\
 \approx 0 & \approx  0  & y^u_3
 \end{pmatrix}
 \label{eq:YuB}
 \end{equation}
Proceeding as in Model A,  the fermion mass hierarchies are 
\begin{equation}
\frac{m_{f_2}}{m_{f_3}} \approx   \frac{ y^{f}_{2 2} }{ y_3^f} \lambda^{\phi}_{f}  \epsilon_\phi\,,
\qquad
\frac{m_{f_1}}{m_{f_2}}\approx  \frac{ y^{f}_{11}  }{ y^{f}_{22} }   y^{\sigma_f}_1     \epsilon_\sigma\,,
\end{equation}
while $3\to2$ quark mixing takes the form  
\begin{equation}
|V_{cb}| \approx  \theta^d_{32} -  \theta^u_{32}  \approx  \left( \frac{y^{d}_Q}{y^d_3} -  \frac{y^{u}_Q}{y^u_3}  \right) y^{\chi_q}_{ 2 }
  \epsilon_\chi\,.
\end{equation}
The $O(\lambda)$ values of  $|V_{td}/V_{ts}|$ and $|V_{ub}/V_{cb}|$, and the related value 
of $|V_{us}|$ 
are controlled by the ratios $y^{u(d)}_{12}/y^{u(d)}_{22} \equiv \theta_{12}^{u,d}$.
In Model B we achieve the minimal breaking of the $U(2)^3$ quark-flavour symmetry proposed  first in~\cite{Barbieri:2011ci}
that, as we shall see next, minimizes the flavour-violating  couplings
of the light fields to the heavy vectors. In this model the down-alignment of the third generation is obtained in the limit $y^{d}_Q \to 0$.

 \begin{table}[t]
 $$\begin{array}{cc|c|c|c|c|c}
& & U(1)_Y^{[3]} & U(1)_{B-L}^{[12]} & U(2)_{T_{3R}}^{[2]}& U(1)_{T_{3R}}^{[1]}& SU(3)\times SU(2)\\ \hline
\multirow{2}{*}{  light VL}   
&  Q  & 1/6 & 0 & 0 &0 &(\bold{3},\bold{2}) \\ \cline{2-7}
&  L  & -1/2 & 0 & 0 &0 &(\bold{1},\bold{2}) \\ \hline 
 \hline
\multirow{3}{*}{ heavy VL   }  
& U & 0& 1/3 & 1/2 &0 &(\bold{3},\bold{1}) \\  \cline{2-7}
& D & 0& 1/3 & -1/2 &0 &(\bold{3},\bold{1}) \\  \cline{2-7}
& E & 0& -1 & -1/2 & 0&(\bold{1},\bold{1}) \\ \hline\hline
{\rm heavy\  Higgs}  &  H^{[12]}_{u,d} & 0 & 0 & -1/2 & 0 &(\bold{1},\bold{2}) \\ \hline 
\end{array}$$
\caption{\small  Non-standard matter content of Model B}
\label{tab:VLFB}
\end{table}

\section{FCNC and bounds on the scales of new physics}
\label{BSNP}

\subsection{Structure of the FCNC couplings}

The lower bounds on the scales of new physics are dominated by tree-level flavour changing neutral current (FCNC) effects
which can be mediated by 
\begin{itemize}
\item  the SM-like $Z$ boson;
\item  the heavier $Z_{3}$ and $Z^\prime_{3}$ bosons, with masses 
$M_{Z_3} \approx g^\prime \langle \phi \rangle, g^\prime \langle \chi \rangle \gg M_Z$,\footnote{Here we generically denote by $g^\prime$ the values of the various  $U(1)$ gauge couplings} coupled universally to the first two (chiral) fermion families via $(B-L)^{[12]}, T_{3R}^{[12]}$ and to the third generation via $Y^{[3]}$;
\item
the heaviest $Z_{12}$ boson, with mass $M_{Z_{12}}\approx g^\prime \langle \sigma \rangle \gg M_{Z_3}$,  coupled non universally to the first two fermion families.
\end{itemize}
 The couplings of these neutral gauge bosons to both chiral fermions ($f$) and VL fermions ($F$), 
 in the interaction basis, can be generically written as 
 \begin{equation}
  J_A^\mu =  g_A \left( \sum_{\psi_L =f_L, F_L} \bar \psi^{(0)}_L \gamma^\mu Q_{\psi_L}^A \psi^{(0)}_L 
  + \sum_{\psi_R =f_R, F_R } \bar \psi^{(0)}_R \gamma^\mu Q_{\psi_R}^A \psi^{(0)}_R \right)~.
 \end{equation}
The effective FCNC interactions are generated by the non commutation of the charges $Q_\psi^A$ with the 
unitary matrices necessary to diagonalise the complete fermion mass matrices.
This mixing can be conveniently  decomposed into: i)~heavy-light mixing, separating heavy and light states,
leading to the effective Yukawa couplings in Eqs.~(\ref{eq:YuA}) and (\ref{eq:YuB}); ii)~light-light mixing necessary to diagonalise the 
effective Yukawa couplings. The distinction between i) and ii) depends on the basis choice adopted for the different 
fields. In the following we adopt the basis giving rise to Eqs.~(\ref{eq:YuA}) and (\ref{eq:YuB}), 
where it is a good approximation to neglect terms 
generated by the product of both heavy-light  and light-light mixing. This allow us to classify the FCNC as follows.
\begin{itemize}
\item {\em FCNC induced by light-light mixing.} \\
To a good accuracy the effective Yukawa couplings in Eqs.~(\ref{eq:YuA}) and (\ref{eq:YuB}) are diagonalised 
by left-handed rotations. In this limit, light-light mixing involves only  Left-Handed (LH) fields. Since $Z_{12}$ is not coupled to LH fields,
and since the $Z$ boson has flavour-universal LH couplings to the light states, only $Z_{3}$ and $Z^\prime_{3}$ are affected.
More precisely, only the combination associated to the $B-L$ generator leads to a FCNC. 
Denoting by $U^f$ the unitary matrices diagonalising on the left side the light Yukawa couplings for a given fermion species, the relevant 
current can be written as 
 \begin{equation}
  J_{3^{(')}}^\mu  \propto  \sum_{f=u,d,e}  \left[ \frac{1}{2} (B-L)_f \right] (U^f_{i3} U^{f*}_{j3})      \bar f^i_L   \gamma^\mu     f^j_L\,. 
  \label{eq:FCNCcurrent}
 \end{equation}
 
 The only case where the small light-light   Right-Handed (RH) mixing 
 plays a non-negligible role is in the $2\to1$ current of the heavy 
 $Z_{12}$ boson, given the non-universal charges, $T_{3R}^{[2]}-T_{3R}^{[1]}$, and the strong experimental bounds in the 1--2 sector.
 Denoting $\tilde U^f$ 
 the unitary matrices diagonalising on the right side the light Yukawa couplings, 
 the relevant  current can be written as 
   \begin{equation}
  J_{12}^\mu  \propto  \sum_{f=u,d,e}  {\tilde U}^f_{12}      \bar f^1_R   \gamma^\mu     f^2_R\,. 
    \label{eq:FCNCcurrent2}
 \end{equation}

\item  {\em  FCNC induced by  heavy-light mixing}\\
Do to the different VL fermion content, the heavy-light mixing is different in the two models and 
we discuss it separately below.
\end{itemize}

\subsubsection{Heavy-light mixing in Model A}
\label{diagModA}

In Model A the complete mass matrices in each sector ($u,d,e$) are $6\times 6$. 
After the symmetry breaking of the full gauge group, the mass matrix of the up-quark sector,
that we analyse in detail  as representative example, has the form
\begin{equation}
{\mathcal L}^{(m_u)}_{\rm eff}  = \begin{pmatrix}
\bar{u}_L^{(0)}&\bar{U}_L^{(0)}
\end{pmatrix}
\begin{pmatrix}
m& \Delta_L \\
\Delta_R &{M}
\end{pmatrix}
\begin{pmatrix}
u_R^{(0)} \\
U_R^{(0)}
\end{pmatrix}\,,
\label{massmatrix}
\end{equation}
where
\begin{equation}
m=v_u
\begin{pmatrix}
0&0&0 \\
0&0&0 \\
0&0& y_3^u
\end{pmatrix},\quad
\Delta_L =v_u
\begin{pmatrix}
y^u_{11}&y^u_{12}  & 0\\
y^u_{21}& y^u_{22}& 0  \\
0&0&0
\end{pmatrix},\quad
\Delta_R =
\begin{pmatrix}
0& y^{\phi_u}_{1 2}  \langle \phi \rangle & 0 \\
0& y^{\phi_u}_{2 2}  \langle \phi \rangle & y^{\chi_u}_{2} \langle \chi^q \rangle  \\
y_1^\sigma  \langle \sigma \rangle  & 0 &   0
\end{pmatrix},
\label{eq:DeltaM}
\end{equation}
all perturbative matrices relative to 
\begin{equation}
M=
\begin{pmatrix}
M_{[23]}&0&  \hat{y}^{\phi_u}_{1 3}  \langle \phi \rangle  \\
0 &M_{[23]}&   \hat{y}^{\phi_u}_{2 3}  \langle \phi \rangle \\
 y^{\phi_u * }_{1 3}  \langle \phi \rangle&  y^{\phi_u * }_{2 3}  \langle \phi \rangle &M_{[12]}
\end{pmatrix}.
\label{eq:BigM}
\end{equation}
The  LH ad RH rotations of the 
$6\times 6$ mass matrix that eliminate the heavy-light mixing can be written as 
 \begin{equation}
 \begin{pmatrix}
u_R^{(0)} \\
U_R^{(0)}
\end{pmatrix} =
\begin{pmatrix}
\bold{1}_R&\epsilon_R\\
-\epsilon^\dagger_R&\hat{\bold{1}}_R
\end{pmatrix}
\begin{pmatrix}
u_R\\
U_R
\end{pmatrix}\,,
\qquad 
 \begin{pmatrix}
u_L^{(0)} \\
U_L^{(0)}
\end{pmatrix} =
\begin{pmatrix}
\bold{1}_L &\epsilon_L\\
-\epsilon_L^\dagger &\hat{\bold{1}}_L
\end{pmatrix}
\begin{pmatrix}
u_L\\
U_L
\end{pmatrix}\,,
\end{equation}
with 
\begin{equation}
\epsilon_R^\dagger \approx M^{-1} \Delta_R \,,\quad 
 \epsilon_L \approx (m\epsilon_R+ \Delta_L ) M^{-1}\,, \quad
\bold{1}_{R(L)} \approx \bold{1} -\frac{1}{2} \epsilon^{\phantom{\dagger}}_{R(L)}\epsilon_{R(L)}^\dagger\,,\quad
\hat{\bold{1}}_{R(L)} \approx \bold{1} -\frac{1}{2} \epsilon^\dagger_{R(L)}\epsilon^{\phantom{\dagger}}_{R(L)}.
\label{3x3matrix}
\end{equation}
where we have neglected terms of $O(\epsilon_{L,R}^3)$.
After applying these rotations, before taking into account light-light mixing, the flavour changing neutral currents in the light sector assume the form
 \begin{equation}
  J_A^\mu =  g_A  \sum_{f_L}  \bar f_L \gamma^\mu \left( 
    \epsilon_L Q_{F_L}^A   \epsilon^\dagger_L  
  - \frac{1}{2} \{ Q_{f_L}^A , \epsilon^{\phantom{\dagger}}_{L}\epsilon_{L}^\dagger  \}  \right) f_L +
 (L \to R)\,.
 \end{equation}
Given the  parametric structure of  $\epsilon_L$ and  $\epsilon_R$, we deduce the following conclusions:
\begin{itemize}
\item 
The LH mixing is very small: it is of $O(v_u/M_{[23]})$ in the up sector, and $O(v_d/M_{[23]})$ 
in the down and charge-lepton sector. This implies that  LH mixing  is completely
negligible for the heavy gauge bosons. It has a relevant phenomenological impact only for the 
effective LH couplings of the $Z$ boson, given that $Q_{f_L}^Z \not= Q_{F_L}^Z=0$. 
These can be written as
  \begin{equation}
  \left. J_Z^\mu \right|_{\rm mod-A} 
   = \sum_{f=u,d,e} \frac{g}{c_W}T_{3L}^f    \bar f^i_L   \gamma^\mu  [\epsilon_L^f  ( \epsilon_L^f)^\dagger]_{ij}  f^j_L,
   \label{eq:JL23}
 \end{equation}
 with
 \begin{equation}
\epsilon_L^f \approx \frac{v_f}{M_{[23]}}
\begin{pmatrix}
y^f_{11}  & y^f_{12}& 0\\
 y^f_{21}& y^f_{22} & 0 \\
0&0 & 0
\end{pmatrix}.
   \end{equation}
So far this is in the flavour basis that leads to Eqs.~(\ref{eq:YuA}) and (\ref{eq:YuB}). Going to the physical basis would introduce  the $U^f$ matrices as in Eq. \ref{eq:FCNCcurrent}. They can however be neglected since all the elements of $\epsilon_L^f$ are of similar order and $U^f\approx 1$.
\item 
The RH mixing is sizable, but it does not 
affect the couplings of the $Z$ boson, since $Q^Z_{f_R} = Q^Z_{F_R} \propto 1$.
The most relevant impact is in the effective couplings of the $Z_3^{(\prime)}$ bosons,
that we can generically write as 
\begin{equation}
  \left. J_{3^{(\prime)}}^\mu \right|_{\rm mod-A} 
   = \sum_{f=u,d,e} g^{(\prime)}_3  [ (\epsilon^f_R)_{2i} (\epsilon^f_R)^*_{3i}  ]    \bar f^2_R   \gamma^\mu     f^3_R + h.c.,\quad i=2,3,
   \label{eq:FCNCcurrent3}
 \end{equation}
with $(\epsilon^f_R)_{2i} \approx O( y^{\phi_f}_{2i} \epsilon_\phi )$ and 
 $(\epsilon^f_R)_{3i} \approx O( y^{\chi_f}_i \epsilon_\chi )$.
 A non-zero $3\to 1$ RH current is generated after taking into account light-light mixing;
 however, this current can be neglected due to the 
 the smallness of RH light-light rotations.
\item 
As already discussed in the case of light-light mixing, 
the heavy $Z_{12}$ boson plays a relevant role only in presence of 
RH mixing in the 1--2 sector. 
 However, the heavy-light RH mixing relevant to the 1--2 sector originates by the matrix $M$ in Eq.~(\ref{eq:BigM})
and not by $\Delta_R$ in (\ref{eq:DeltaM}). This implies that it is strongly suppressed:
 $(\epsilon^u_R)_{12}  = O( \langle \phi \rangle /M_{[12]}) =  O(\epsilon_\phi M_{[23]} /M_{[12]})  \ll  O(\epsilon_{\chi,\phi})$. 
 Hence the  FCNC couplings of the $Z_{12}$ resulting from 
heavy-light mixing  are subleading.
\end{itemize}

\subsubsection{Heavy-light mixing in Model B}
\label{diagModB}

In Model B the complete mass matrices in each sector ($u,d,e$) are $5\times 5$. 
After the symmetry breaking of the full gauge group, the mass matrix of the up-quark sector,
again  as representative example, has the same form as eq. (\ref{massmatrix}) where now $U_{L,R}$ are bi-dimensional vectors
\begin{equation}
U=
\begin{pmatrix}
Q_u\\
U
\end{pmatrix},
\end{equation}
 and
\begin{equation}
m=v_u
\begin{pmatrix}
0&y^u_{12} \lambda_u^\phi \epsilon_\phi&0 \\
0&y^u_{22} \lambda_u^\phi \epsilon_\phi&0 \\
0&0& y_3^u
\end{pmatrix},\quad
\Delta_L =
\begin{pmatrix}
0&y^u_{11} \lambda_u^\phi \epsilon_\phi v_u \\
y_2^{\chi_q}\langle \chi \rangle&y^u_{21} \lambda_u^\phi \epsilon_\phi v_u \\
0&0
\end{pmatrix},\quad
\Delta_R =
\begin{pmatrix}
0  & 0 &   y_Q^u v^u  \\
y_1^{\sigma_u}  \langle \sigma \rangle& 0&0  
\end{pmatrix},
\label{eq:DeltaMB}
\end{equation}
all perturbative matrices relative to 
\begin{equation}
M=
\begin{pmatrix}
M_Q&0 \\
0&M_{[12]}
\end{pmatrix}.
\label{eq:BigMB}
\end{equation}

Proceeding as in the case of Model A, one sees that there is no flavour changing interaction 
in the RH currents\footnote{The heavy-light RH mixing leads to a breaking of universality 
of the RH coupling of the $Z$ boson, among light generations, of $O[(m_{f_1}/m_{f_2})^2]$.
This effect is below current experimental errors. If combined with RH mixing in the light-light sector, 
this would in turn induce flavour-violating effects, such as 
 $\cB(\mu \to 3e) \approx 5 \times 10^{-13} |(\theta^e_R)_{12}/10^{-2}|^2$.  }
 and the possible effects due to  LH mixing in the $Z_3^{(\prime)}$ currents are either vanishing due to $(\epsilon^f_L)_{11}=(\epsilon^f_L)_{31}=0$, 
or negligibly small due to $(\epsilon^f_L)_{21}\approx (\epsilon^f_L)_{22}\approx \epsilon_\phi v_f/M_{[12]}$.
We recall that the condition $(\epsilon^f_L)_{11}=(\epsilon^f_L)_{31}=0$ is due to the specific basis choice we have adopted for the $q_i$ fields. 
With this basis choice we  ``move" all the potential FCNC effects arising from the  heavy-light mixing into the light-light sector.

\subsection{Experimental bounds}

\subsubsection{Constraints from tree-level FCNC amplitudes}

Using the effective  interactions discussed in the previous section, we can derive a series of bounds from $\Delta F=2$ and $\Delta F=1$ 
FCNC processes. The results, summarised in Table~\ref{TableBounds}, can be described as follows.

 \begin{table}[t]
 \begin{center}
\begin{tabular}{c|c|c||c|l}
Model & Mixing  &  Mediator & Observable & Bound (TeV)   \\ 
\hline \multirow{3}{*}{A,B}  & \multirow{2}{*}{$({\rm light-light})_{\rm LH}$}  & \multirow{2}{*}{$Z_3^{(\prime)}$}  
  		& $\Delta B_{d,s}=2$ 	& $ \langle\chi\rangle  >   2.7 \, |\theta_{23}^d|/|V_{cb}|$    \\[1pt]  
  &  & 	& $B_s\to\mu^+\mu^-$ 	& $ \langle\chi\rangle  >   5.6\,  \sqrt{ |\theta_{23}^d|/|V_{cb} |}$   \\[2pt] \cline{2-5}  
  &  $({\rm light-light})_{\rm RH}$  &   $Z_{12}$   	
  		& $\Delta S=2$		& $ \langle\sigma\rangle   > 280\,  | {\rm Im}(\theta^d_R)_{12}|/10^{-2}  \phantom{\Big]}  $  \\[2pt]
\hline\hline  \multirow{6}{*}{A}  & \multirow{3}{*}{$({\rm heavy-light})_{\rm LH}$}  & \multirow{3}{*}{ $Z$}
 				& $\Delta C=2 $		& $M_{U_{1,2}}  >  30 \times O(|y^u_{ij}|)$ \\ 
&  & 				& $\Delta S=2 $		& $M_{D_{1,2}}  >  50 \cot\beta \times O(|y^d_{ij}|) $ \\ 
&  & 				&$\mu \rightarrow 3e$ 	& $M_{E_{1,2}}  >    250 \cot\beta \times O(|y^e_{ij}|) $
\\[2pt]  \cline{2-5}     & \multirow{3}{*}{ $({\rm heavy-light})_{\rm RH}$ }  & \multirow{3}{*}{ $Z_3^{(\prime)}$ }
				& $\Delta B_s=2$		& $M_{D_{1,2}}  > 14 \sqrt{ \langle \phi \rangle /{\rm TeV}} \times  O(|y^{\phi_d, \chi_d}_{ij} |) $   \phantom{\Big]}   \\[2pt] 
&  & 				& $B_s\to\mu^+\mu^-$      & $M_{D_{1,2}}  > 25 \times O(|y^{\phi_d, \chi_d}_{ij} |) $ \\[2pt] 
&  & 				& $\tau\rightarrow 3\mu$	& $M_{E_{1,2}} > 20 \times O(|y^{\phi_e, \chi_e}_{ij} | ) $ 

\end{tabular} 
\end{center}
\caption{\small  Bounds in TeV on the symmetry-breaking scales of the gauge group ($\langle \chi^{q,l} \rangle \approx \langle \phi \rangle$ and $\langle \sigma \rangle$)
and on the VL fermion masses  ($M_{F_\alpha}$). See text for more details.
}
\label{TableBounds}
\end{table}

\begin{itemize}
\item{}
The most precise bounds are those following from light-light LH mixing that, in first approximation, 
depend only on the parameter $\theta^d_{32}$ controlling the diagonalization of the 
down Yukawa couplings (in both models).
The effective interaction following from the tree-level $Z_3^{(\prime)}$ exchange can be written as
\begin{equation}
\mathcal{L}_{\rm eff}^{\rm FCNC} = \frac{ 1  }{ \langle \chi \rangle^2  } J_{3^{(\prime)}}^\mu J_{{3^{(\prime)}} \mu}\,,
\label{eq:LFCNC}
\end{equation}
where $\langle \chi \rangle \equiv  \langle \chi^{q,l } \rangle \approx  \langle \phi \rangle $ and  $J_{3^{(\prime)}}^\mu$ contains both the FCNC term in~(\ref{eq:FCNCcurrent}) as well as the flavour-conserving terms 
 associated to the $Z_3^{(\prime)}$ generators.\footnote{The normalization in  Eq.~(\ref{eq:LFCNC})
 is such that the r.h.s. of Eq.~(\ref{eq:FCNCcurrent}) appears in  $J_3^\mu$ without extra coefficients, 
 in the limit $g^{[3]} \gg g^{[12]}$ and defining $\langle \chi \rangle$ via $M_{Z_3} = g^{[3]}  \langle \chi \rangle/\sqrt{2}$.}
 The four-fermion operators thus obtained are in one-to-one correspondence with those recently  analysed in~\cite{Allwicher:2023shc}, 
 whose results we use to derive numerical bounds.  
The square of the FCNC term leads to a $\Delta B=2$ effective operator ($Q^{(1)[3333]}_{qq}$ in the notation of \cite{Allwicher:2023shc})
 whose bound is not particularly stringent due to the smallness of the $B$-$L$ charge of the quarks (as noted first in~\cite{Allanach:2018lvl} in a similar context).
  \item{} 
The most stringent bound on light-light LH mixing arises, at present, from $B_s\to\mu^+\mu^-$. The amplitude for this  process 
receives two contributions, with different parametric dependence on the different gauge couplings. 
A first contribution is due to the $Z_3^{(\prime)}$ exchange among two fermion currents: 
this contribution is suppressed as $(g^{[12]} / g^{[3]})^2$ in the large  $g^{[3]}$ limit. 
As noted in~\cite{Davighi:2023evx}, an independent contribution  
arises by the $Z_3^{(\prime)}$--$Z$ mixing,  which occurs after the breaking of the SM electroweak symmetry. 
In the SM-EFT language, this contribution is described by the effective operator  $Q^{(1)[33]}_{Hq}$ (in the notation of \cite{Allwicher:2023shc}), 
which appears thanks to the Higgs current in $J_{3^{(\prime)}}^\mu$.
Note that $Z_3^{(\prime)}$--$Z$ mixing is unavoidable given Higgs, $\chi^{q,l }$,  and $\phi$ fields are all charged under $U(1)_Y^{[3]}$.
Moreover, while this mixing vanishes in the limit $v_{u,d}/\langle \phi,\chi \rangle \to 0$, the corresponding contribution to 
the $B_s\to\mu^+\mu^-$ amplitude, being further proportional to $g^2_Z/m_Z^2 \sim 1/v^2$, does not.   
In the limit $g^{[3]} \gg g^{[12]}$, which we adopt to derive the numerical entries in the first two rows in Table~\ref{TableBounds},
the $B_s\to\mu^+\mu^-$ amplitude is dominated by the $Z_3^{(\prime)}$--$Z$ mixing contribution.
\item{} Proceeding in a similar manner we derive the bounds from light-light RH mixing, which also apply to both models and 
which depend only on the RH mixing among the first two generations. We denote the latter $(\theta^d_R)_{12}$. 
The experimental constraint on the $\Delta S=2$ operator is taken from~\cite{Bona:2022zhn}.
\item{} 
The bounds derived from heavy-light mixing, which apply only to Model A, involve more model parameters. 
The numerical values reported in Table~\ref{TableBounds} correspond to assuming $O(1)$ values for all the 
Yukwa couplings in Eq.~(\ref{eq:YuA}). However, they all scale linearly with a given set of Yukawa couplings,
as explicitly indicated.  For instance the bound on VL masses from $\Delta C=2$ decreases from 30~TeV to 3~TeV 
for $y^u_{ij}\approx 1 \to 0.1$.  Note the appearance of $\cot\beta \approx 0.02 \div 0.03$ on the $Z$-mediated 
amplitudes in the down-quark and charged-lepton sector. 
\end{itemize}
 The bounds reported in  Table~\ref{TableBounds} correspond to the present state of the art of 
 flavour-physics measurements.  It is interesting to stress that many of the corresponding entries will improve significantly in the next 
 $10\div 15$ years. Just to mention two examples: from the expected improvements on CKM inputs (see e.g.~\cite{LHCb:2018roe}), 
 combined with Lattice-QCD 
 estimates of the $\Delta B=2$ hadronic operators,  we expect the entry on the first row to improve by about 
 a factor of $\approx 3$ (in absence of deviations from the SM). A similar level of improvement can be expected on clean semileptonic 
 FCNC transitions, hence on the second row of the table, taking into account the projected errors on 
 $\cB(B_s\to\mu^+\mu^-)$~\cite{LHCb:2018roe}, and the clean neutrino modes $\cB(K^+\to\pi^+\nu \bar \nu)$~\cite{HIKE:2022qra} and 
 $\cB(B\to K\nu \bar \nu)$~\cite{Belle-II:2018jsg}.

\subsubsection{FCNCs beyond the tree level}
Additional FCNC amplitudes  are generated beyond the tree level
from one-loop diagrams with internal heavy $Z^{(\prime)}$ bosons and 
light or heavy fermions. These amplitudes can be of two types: 1) four-fermion structures already present at the tree level, 2) new electroweak structures, 
such as dipole operators,  which are generated only beyond the tree level. A detailed analysis of all one-loop amplitudes 
is beyond the scope of this work. In the following we limit ourself to point out the potentially largest 
contributions. In particular, we highlight constraints exceeding, or parametrically different
with respect to those derived from  the tree-level mediated FCNC ampltides discussed above.

\paragraph{Four-fermion structures}  We first note that FCNC amplitudes are a result of the Yukawa interactions, 
which are the only interactions responsible for flavour violations. This implies that, to a good approximation, the leading 
 four-fermion FCNC amplitudes can be more efficiently computed in the gauge-less limit of the theory.  
 
The $\Delta F=1$ amplitudes generated beyond the tree level, with external light fermions, 
necessary involve the product of two FCNC vertices, e.g.  $\mathcal{V}_{H}(f_i \to F_k)  \times  \mathcal{V}_{H}(F_k \to f_j)$,
where $H$ denotes any of the the scalar bosons of the theory (including Goldstone bosons in the gauge-less limit).
The potentially more interesting case are FCNC transitions of the type $2\to 1$,  with an intermediate heavy fermion: 
here the two  vertices  $\mathcal{V}_{H}(f_i \to F_k)$ and   $\mathcal{V}_{H}(F_k \to f_j)$  can be significantly larger 
than the tree-level  effective coupling $\mathcal{V}_{Z^{(\prime)}}(f_2 \to f_1)$, and overcome the loop suppression. 
The same reasoning applies to $\Delta F=2$ transitions, with the only difference that in this case loop amplitudes involve four FCNC vertices.

In Model A the amplitude leading to the most stringent constraint is the  $\Delta S=2$ amplitude resulting from the 
box with heavy fermions and the (SM-like) $H_d$ field. This leads to the following effective operator 
(in the gauge-less limit, and assuming $M_{D_\alpha} \gg M_{H_d}$):
\begin{equation}
 \frac{ ( y^d_{2\alpha} y^{d*}_{1\alpha})^2 } {64\pi^2 (M_{D_\alpha})^2 }  (\bar s_L \gamma^\mu d_L )^2  
 \label{eq:DS2loopA}
\end{equation}
From the constraint on the $\Delta S=2$ operator in~\cite{Bona:2022zhn} we deduce
\begin{equation}
M_{D_2} \approx M_{[23]}   > 180~{\rm TeV} \times \frac{ \sqrt{ |{\rm Im}  (y^d_{22} y^{d*}_{12})^2 |} }{ |V_{us}| }\,,
 \label{eq:DS2loopA2}
\end{equation}
and similarly for  $M_{D_1}$. This limit is significantly stronger that those reported in Table~\ref{TableBounds}  for 
$M_{D_{1,2}}$, unless the CP-violating phase of the relevant Yukawa combination is unnaturally small. 

In Model B all contributions associated to the $\Lambda_{[23]}$ scale are protected by the minimal breaking of  
$U(2)^5$ occurring at that scale: the $\Delta S=2$ loop amplitudes are proportional to $ ( V_{td} V_{ts}^*)^2$ and are subleading compared to the tree-level ones. A different combination of flavour-violating couplings appears only when the heavy $SU(2)_L$--singlet fermions associated 
to the $\Lambda_{[12]}$ scale are involved. The most stringent constraint follows from the $\Delta S=2$ box induced by 
the $D$--$H_{[12]}$ exchange. This lead to the following effective operator 
(in the gauge-less limit, and assuming $M_D \gg M_{H_{[12]}}$):
\begin{equation}
 \frac{ ( y^d_{21} y^{d*}_{11})^2 } {64\pi^2 M_D^2 }  (\bar s_L \gamma^\mu d_L )^2\,,
 \label{eq:DS2loopB}
\end{equation}
which implies
\begin{equation}
M_{D} \approx M_{[12]}   > 80~{\rm TeV} \times \sqrt{{\rm Im}\left[ 
 \left( \frac{ (\theta_R^d)_{12} }{ 0.1\, \epsilon_\sigma }  \frac{ y^d_{22} y^{d*}_{11} }{ y_1^\sigma}  \right)^2\, \right]} \,.
 \label{eq:DS2loopB2}
\end{equation}
This limit is comparable to that on $\langle \sigma \rangle$ in Table~\ref{TableBounds}: it provides a further indication that the 
$\Lambda_{[12]}$ scale is at least of $O(100)$~TeV,  and that its lower allowed value is controlled 
by the right-handed mixing in the $1$--$2$ sector. 
It is worth stressing that  Eq.~(\ref{eq:DS2loopA2}) and Eq.~(\ref{eq:DS2loopB2}), 
which are similar from a numerical point of view, have very different implications on the respective models 
since they  refer to the low scale ($\Lambda_{[23]}$) in model B and to the high scale ($\Lambda_{[12]}$) in model A.

\paragraph{FCNC dipole amplitudes} In this class of amplitudes the most stringent constraint is represented by $\mu \to e \gamma$,
given its strong experimental bound~\cite{MEGII:2023ltw}. At variance with respect to four-fermion amplitudes, in this case it is 
easier to evaluate the leading contributions via the gauge-boson exchange in the unitary gauge. 

In Model A we can identify two contributions: those where the $Z_3^{(\prime)}$  has both a LH and a RH vertex, and those with two RH vertices.
The RH$\times$RH one is necessarily suppressed by the electron Yukawa coupling. The potentially larger  
LH$\times$RH contribution due to the $E_{\alpha}$--$Z_3^{(\prime)}$ loop, 
 computed in the limit $M_{E_{\alpha}} \gg M_{Z^\prime}$, 
leads to the following  effective operator 
\begin{equation}
  \frac{(g^{[3]})^2}  {64\pi^2} \frac{M_{E_\alpha} }{ M_{Z^\prime}^2  }  (\epsilon_L^{i\alpha} \epsilon^{\dagger \alpha 2}_R)\,  \bar \ell^i_L \sigma^{\mu\nu} (eF_{\mu\nu})  \mu_R
\end{equation}
with the $\epsilon_{L,R}$ defined as in Section~\ref{diagModA}.
However, the above result holds before considering light-light mixing, i.e.~the diagonalization of the analog of Eq.~(\ref{eq:YuA}) for charged leptons. 
The mixing term has the following form
\begin{equation}
 \epsilon_L^{i\alpha} \epsilon^{\dagger \alpha 2}_R \approx \frac{ m_\mu}{ M_{[23]}} \left[   \frac{ y^e_{i \alpha} y^{\phi_e}_{\alpha 2} }{ y^e_{2 \alpha} y^{\phi_e}_{\alpha 2} }
 + ( y^{\phi_e}_{\alpha 2} )^2 \epsilon_\phi^2 \right]~.
\end{equation}
Note that the first term is fully aligned with the Yukawa coupling (independently of the approximations so far employed) and becomes 
flavour diagonal in the mass mass-eigenstate basis. The surviving flavour-violating effect leads  leads to 
\begin{equation}
 \frac{1} {32\pi^2} \frac{m_\mu  }{ M^2_{[23]} }  [(y^{\phi_e}_{\alpha 2} )^2  U_{12}^e]\,  \bar e_L \sigma^{\mu\nu} (eF_{\mu\nu})  \mu_R~.
 \label{eq:mueg}
\end{equation}
Contributions of comparable size (and similar parametric form) are obtained when considering the mass splitting among the
vector-like fermions.  Given the present experimental limit on $\mu \to e \gamma$~\cite{MEGII:2023ltw}, 
the result  (\ref{eq:mueg}) translates into the bound 
\begin{equation} 
M_{[23]} > 12~{\rm TeV} \sqrt{ (y^{\phi_e}_{\alpha 2} )^2 U_{12}^e / 10^{-2} }~,
\end{equation}
which is comparable to those reported in Table~\ref{TableBounds} for Model A.

In Model B, similarly to the case of four-fermion operators, the potentially dangerous contributions arise only by loop diagrams 
involving the heaviest VL fermion. The leading LH$\times$RH contribution
 due to the $E$--$Z_3^{(\prime)}$ loop,  computed in the limit $M_{E} \gg M_{Z^\prime}$, 
give rise  to the following  effective operator 
\begin{equation}
  \frac{(g^{[3]})^2}  {64\pi^2} \frac{M_{E} }{ M_{Z^\prime}^2  }  (\epsilon_L^{2 3} \epsilon^{\dagger 3 1}_R)\,  \bar \mu_L \sigma^{\mu\nu} (eF_{\mu\nu})  e_R
  \approx \frac{ 1 }{32\pi^2}  \frac{m_\mu  }{ \langle \phi \rangle^2 } \, (\theta^l_R)_{12}\, \bar \mu_L \sigma^{\mu\nu} (eF_{\mu\nu})  e_R~.
  \label{eq:muegB}
\end{equation}
with the $\epsilon_{L,R}$ defined as in Section~\ref{diagModB}.The result on the r.h.s.~follows from expressing the effective coupling in terms of the charged-lepton 
Yukawa coupling, and defining $(\theta^e_R)_{12} =  Y^e_{21}/Y^e_{22}$
(note that in this case the result is essentially unaffected by light-light mixing).
The corresponding bound can be written as
\begin{equation} 
\langle \phi \rangle  > 12~{\rm TeV} \sqrt{ (\theta^e_R)_{12} / 10^{-2} }~.
\end{equation}
This  is the strongest constraint on $\langle \phi \rangle$ in Model B
unless $|(\theta^e_R)_{12} | \lsim 10^{-3}$,  below the natural size $|(\theta^e_R)_{12} | \sim m_e/m_\mu$.

\subsubsection{Bounds from electroweak precision observables and direct searches}
From the previous analyses we conclude that, at least in Model B, the $Z_3^{(\prime)}$ bosons  
can satisfy the flavour bounds with masses as low as $2\div3$~TeV, assuming a modest  down-alignment
and $|(\theta^e_R)_{12} | \lsim 10^{-3}$.
For such low masses, electroweak precision observables (EWPO) and direct searches become   relevant.
An extensive study of such bounds for the $Z^\prime$ associated to the 
$U(1)_Y^{[3]} \times U(1)_Y^{[12]}$ breaking has been presented in~\cite{Davighi:2023evx}.
In that context, it has been shown that the combined bounds from direct searches (via $pp\to \ell^+\ell^-$)
and EWPO (in particular the correction to $m_W/m_Z$) require $M_{Z^\prime} >  4.5$~TeV.
The same conclusion applies in our case (up to minor modifications): the strong bounds from $pp\to \ell^+\ell^-$, for $\ell=e,\mu$, 
can be reduced in the limit $g^{[3]} \gg g^{[12]}_{B-L},  g^{[12]}_{R}$; however, 
the $Z_3^{(\prime)}$--$Z$ mixing, hence electroweak constraints, are maximised in this limit. 

It is worth stressing that the difficulty of evading both direct searches and electroweak constraints is dictated by the 
non-vanishing charge of the $Z_3^{(\prime)}$ under $U(1)_Y^{[3]}$, which necessarily implies 
a non-vanishing $Z_3^{(\prime)}$--$Z$ mixing. In the extended models discussed in~\cite{Davighi:2023iks}, 
where $U(1)_Y^{[3]}$ is further decomposed into $U(1)_{B-L}^{[3]}$ and
$U(1)_{T_R}^{[3]}$, the $Z^\prime$ boson associated to $B-L$ does not mix
at the tree level with the SM-like $Z$ boson, hence it can have a mass as low as $2\div 3$~TeV
in in the limit $g^{[3]}_{B-L} \gg g^{[12]}_{B-L}$.

\section{Extension to neutrinos}
\label{ETN}

To describe neutrino masses and mixings, we extend Models A and B  as shown in Table \ref{VLFneutrinos}, including  two scalars, $\sigma_i$
($i=1,2$),  and an appropriate  set of vector-like neutrinos: $\{N_{\alpha},  N_3\}$ in Model A and only $N$ in Model B.
In both cases the new fermions complete the families of vector-like singlets under $SU(3)\times SU(2)$ with a state which is a total singlet under the SM gauge group. We recall that, by construction, both models have three families of chiral right-handed neutrinos: the $U(1)$ charges of
 $\nu_R^{1,2}$  are fixed in order to cancel gauge anomalies,  while $\nu_R^3$ is a singlet of the full gauge group.

\begin{figure}[t]
\centering
\includegraphics[clip,width=.85\textwidth]{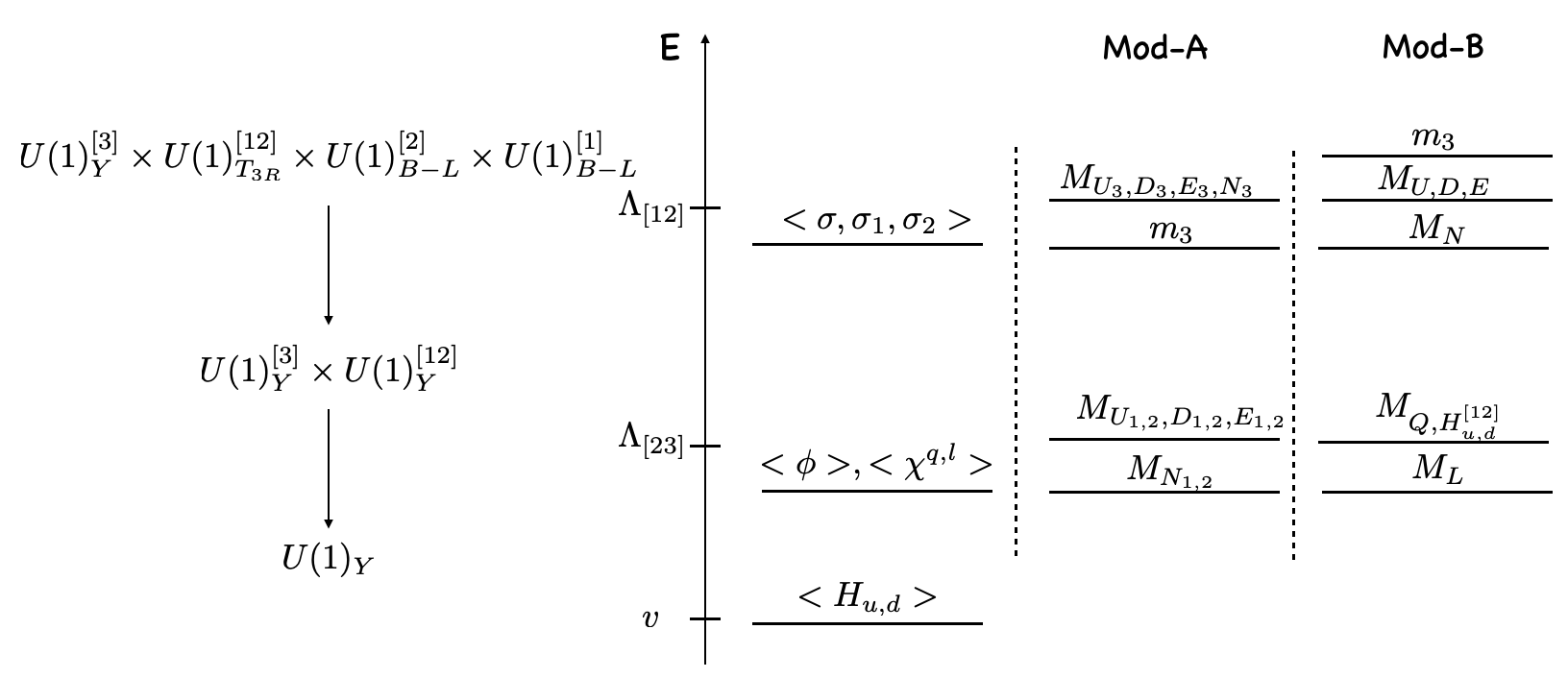}
\caption{\small  The three scale picture  of Models A and B with the inclusion of the additional fields need to describe neutrino masses}
\label{fig:Fig-ABnu}
\end{figure}

All the gauge invariant Yukawa-like and fermion mass terms are obtained by adding the lepton-number  violating terms (with proper Lorentz contractions left understood)
\begin{equation}
\mathcal{L}^\nu_{L-{\rm breaking}} =  
 y^{\sigma_1}\sigma_1\nu_R^1\nu_R^1 +  y^{\sigma_2} \sigma_2\nu_R^2\nu_R^2 + m_3 \nu_R^3\nu_R^3\,,
\label{Lnu}
\end{equation}
to Eqs.~(\ref{LYA}) and  (\ref{LYB}) in Model A and B respectively, with the $q,u,U$--fields replaced by  $l,\nu,N$.
The $\sigma_i$ fields, interacting with $\sigma$ via the $d=4$ coupling $\sigma_1^*\sigma_2\sigma^2$, get their vevs at the same scale of $ \langle \sigma \rangle$, such that
\begin{equation}
U(1)_Y^{[3]} \times U(1)_{B-L}^{[12]} \times U(1)_{T_{3R}}^{[2]} \times U(1)_{T_{3R}}^{[1]} 
\stackrel{\langle\sigma,\sigma_1,\sigma_2\rangle}{\longrightarrow}
U(1)_Y^{[3]} \times U(1)_{Y}^{[12]} 
\stackrel{\langle\phi,\chi\rangle}{\longrightarrow}
 U(1)_Y~.
\end{equation}
In Fig.~\ref{fig:Fig-ABnu} we indicate the reference energy scale, and corresponding splitting, for the masses of the new fields introduced in Table \ref{VLFneutrinos}. Note the displacement, relative to Fig.~\ref{fig:Fig4overall}, of $M_L$, which gives $\epsilon_\chi\approx 1$, thus affecting the mixing angle in the charged lepton sector, $\theta^l_{23}\approx 1$, but not $m_\mu/m_\tau \propto \epsilon_\phi$.

\begin{table}[t]
 $$\begin{array}{cc|c|c|c|c|c}
& & U(1)_Y^{[3]} & U(1)_{B-L}^{[12]} & U(2)_{T_{3R}}^{[2]}& U(1)_{T_{3R}}^{[1]}& SU(3)\times SU(2)\\ \hline
{\rm light\  VL} & N_{1,2} & 1/2& -1 & 0 &0 &(\bold{1},\bold{1}) \\ \hline\hline
{\rm heavy\  VL} &N_3 & 0& -1 & 0 &1/2 &(\bold{1},\bold{1}) \\  \hline\hline
\multirow{2}{*}{scalars}
&\sigma_1 & 0& 2 & 0 & -1&(\bold{1},\bold{1}) \\ \cline{2-7}
&\sigma_2 & 0& 2 & -1 & 0&(\bold{1},\bold{1}) \\ \hline
\end{array}$$
\caption{\small   Quantum numbers of the additional scalars ($\sigma_{1,2}$) and VL fermions ($N_{\alpha},  N_3$) added 
to the two models to describe neutrino masses and mixings.
In Model B only the VL fermion $N_3\equiv N$ is present. }
\label{VLFneutrinos}
\end{table}

In Model A, in analogy with $Y_u$, the neutrino Yukawa coupling matrix $Y_{\nu A}$ has a similar form to Eq.~(\ref{eq:YuA}), except for $\epsilon_\phi\approx\epsilon_\chi\approx 1$ due to $M_{N_\alpha} \approx \langle\phi\rangle$. This way the Majorana mass matrix of the left-handed neutrinos, after integrating out the right-handed neutrinos in (\ref{Lnu}), assumes the form
\begin{equation}
 m^\nu_{LL} |_{\rm Mod.~A}  \approx v_u^2 \begin{pmatrix}
  \frac{y^4}{y^{\sigma_2} \langle \sigma_2 \rangle} + \frac{y^4}{m_3} &  \frac{y^4}{y^{\sigma_2} \langle \sigma_2 \rangle} + \frac{y^4}{m_3} & \frac{y^5}{y^{\sigma_2} \langle\sigma_2 \rangle} + \frac{y^3}{m_3} \\
 \frac{y^4}{y^{\sigma_2} \langle \sigma_2 \rangle} + \frac{y^4}{m_3} &  \frac{y^4}{y^{\sigma_2} \langle \sigma_2 \rangle} + \frac{y^4}{m_3} & \frac{y^5}{y^{\sigma_2} \langle\sigma_2 \rangle} + \frac{y^3}{m_3} \\
 \frac{y^5}{y^{\sigma_2} \langle \sigma_2\rangle} + \frac{y^3}{m_3} &  \frac{y^5}{y^{\sigma_2} \langle\sigma_2\rangle} + \frac{y^3}{m_3} & \frac{y^5}{y^{\sigma_2} \langle \sigma_2\rangle} + \frac{y^2}{m_3} 
 \end{pmatrix},
 \label{eq:MLLA}
 \end{equation}
where terms of order $\epsilon_\sigma^2 y^n/(y^{\sigma_1} \langle \sigma_1 \rangle) $ have been neglected. To simplify the notation, we have 
indicated only the overall power of the dimensionless couplings $y$, all different from each other, that appear in every term. To this end one has to include as well, from the analogue of Eq. \ref{eq:YuA}, the entry $(Y_{\nu A})_{32}\approx y_3^\nu y_3^{\chi_\nu} y_{32}^{\phi_\nu}$.

In a similar way, in Model B the neutrino Yukawa coupling matrix $Y_{\nu B}$ has the form of (\ref{eq:YuB}) except for $\epsilon_\sigma\approx 1$, due to  $M_{N_1}\approx  \langle\sigma\rangle$, and $\epsilon_\chi\approx 1$, due to  $M_L\approx  \langle\phi\rangle$. 
Integrating out the heavy states the leads to 
\begin{equation}
 m^\nu_{LL} |_{\rm Mod.~B}  \approx v_u^2 \begin{pmatrix}
  \epsilon_\phi^2(\frac{y^6}{y^{\sigma_1} \langle \sigma_1 \rangle} + \frac{y^4}{y^{\sigma_2} \langle \sigma_2 \rangle}) &  
 \epsilon_\phi^2(\frac{y^6}{y^{\sigma_1} \langle \sigma_1 \rangle} + \frac{y^4}{y^{\sigma_2} \langle \sigma_2 \rangle}) & 0 \\
 \epsilon_\phi^2(\frac{y^6}{y^{\sigma_1} \langle \sigma_1 \rangle} + \frac{y^4}{y^{\sigma_2} \langle \sigma_2 \rangle}) &  
 \epsilon_\phi^2(\frac{y^6}{y^{\sigma_1} \langle \sigma_1 \rangle} + \frac{y^4}{y^{\sigma_2} \langle \sigma_2 \rangle}) +  \frac{y^4}{m_3} &  \frac{y^3}{m_3}  \\
0 &   \frac{y^3}{m_3} &  \frac{y^2}{m_3} 
 \end{pmatrix},
 \label{eq:MLLB}
 \end{equation}

The mass matrices in Eq.~(\ref{eq:MLLA}) and (\ref{eq:MLLB}) provide a good  phenomenological 
description of neutrino masses and mixings under two main conditions: a relation between  
$m_3$ and $\langle \sigma \rangle$, and  an appropriate overall normalization. 
The first condition for the two models is (See Fig.~\ref{fig:Fig-ABnu} )
\begin{equation}
m_3\approx \left\{    
	\begin{array} {ll}  
	\langle \sigma \rangle & {\rm Model~A}\,,  \\
	\langle \sigma \rangle/\epsilon_\phi^2 \qquad\qquad & {\rm Model~B}\,.
	\end{array}
\right.
\end{equation}
Under this conditions all the dimensionless entries in Eq.~(\ref{eq:MLLA}) and (\ref{eq:MLLB}) 
can vary in a range $0.1\div 1$ up to their overall normalization. 
The normalization condition, necessary to achieve $m_\nu\approx 0.1$~eV for the 
largest neutrino mass eigenvalue, is 
\begin{equation}
\langle \sigma \rangle  \approx \left\{    
	\begin{array} {ll}  
	(y^\nu_H)^2\times 10^{14\div 15}~{\rm GeV}  & {\rm Model~A}\,,  \\
	\epsilon_\phi^2(y^\nu_H)^2\times 10^{14\div 15}~{\rm GeV} \qquad\qquad & {\rm Model~B}\,.
	\end{array}
\right.
	\label{eq:nu-cond}
\end{equation}
In these expressions we have factored out the universal dependence of all entries in Eqs.~(\ref{eq:MLLA}) and (\ref{eq:MLLB}) on any Yukawa coupling  ($y^\nu_H$) which involves the Higgs fields  in Eqs.~(\ref{LYA}) and (\ref{LYB}). 
Satisfying the condition (\ref{eq:nu-cond}) requires either a very large value of $\langle \sigma \rangle$ in both cases or, to saturate the lower bound on $\langle \sigma \rangle$  from flavour observables, $y^\nu_H\approx 10^{-5}$ in Model A and $y^\nu_H\approx 10^{-(3\div 4)}$ in Model B.

\medskip 

Assuming  the $d=4$ coupling $\sigma_2 \sigma_1^*\sigma^2$ is present in scalar potential, the first stage of breaking of the gauge group,
$U(1)_Y^{[3]} \times U(1)_{B-L}^{[12]} \times U(1)_{T_{3R}}^{[2]} \times U(1)_{T_{3R}}^{[1]} \to U(1)_Y^{[3]} \times U(1)_{Y}^{[12]}$, does not lead to any massless boson. On the contrary, a peudo-Goldstone boson appears in the second stage of breaking, 
$U(1)_Y^{[3]} \times U(1)_{Y}^{[12]} \to U(1)_Y$. This state  receives mass from a $d=5$ scalar potential term $(\phi^* \chi^l)^2\sigma_2$. Such a term,  in fact generated at one loop at the scale 
$\langle \sigma \rangle$, gives a mass $m_G\approx 0.1 \langle \phi \rangle$ to the pseudo-Goldstone boson.

\section{Summary and conclusions}

We have presented a complete description of flavour which does not invoke small dimensionless couplings, 
but follows from the hypothesis of a multi-scale flavour non-universal UV extension of the SM gauge group. The main idea of the construction 
is summarised in Fig.~\ref{fig:Fig4overall}, where we illustrate the two-step symmetry-breaking chain from the non-universal 
gauge group in Eq.~(\ref{eq:G}) down to the SM. As a consequence of this breaking chain,  the SM Yukawa couplings 
assume the parametric form in Eq.~(\ref{eq:Ypar}), which describes well masses and mixing angles of quarks and charged leptons 
in terms of $O(1)$ couplings and two scale ratios. Within this general framework, rather than sticking to an EFT-level description, 
we have constructed and analysed two concrete renormalizable models with different field content, also summarised in Fig.~\ref{fig:Fig4overall}, 
where the three SM Yukawa coupling matrices are fully calculable, although in terms of many unknown couplings
of size $0.1\div 1$.

The construction in Fig.~\ref{fig:Fig4overall}, can be viewed as the minimal extension of the SM 
 yielding a phenomenologically viable (and calculable) structure for the Yukawa coupling matrices in absence of small couplings. 
The apparently more economical choice based on a flavour-deconstruction of hypercharge only, discussed in the Appendix,
requires either small couplings or a rather cumbersome field content.   
The two models we have considered can also describe neutrino masses, with an approximate anarchic spectrum, 
 via the minimal extensions illustrated in Fig.~\ref{fig:Fig-ABnu}. 
 In this case, the only price to pay to be consistent with data is an  adjustment of the  overall  normalization of the neutrino  mass matrix.  

Having constructed two explicit models, we have been able to investigate in detail the constraints on the exotic fields and, 
more generally, on the UV scales $\Lambda_{[23]}$ and $\Lambda_{[12]}$, dictated by present data on flavour-changing processes. 
As we have show, at least in Model~B the scale $\Lambda_{[23]}$ can be as low as 10 TeV, i.e.~some of the heavy $Z^{(\prime)}$ bosons, of $O({\rm TeV})$ mass,
could be close to present bounds from direct searches. In this case, constraints from EWPO, direct searches, and flavour-changing processes
 appear to be, at the present time, of comparable importance (as noted also in~\cite{Davighi:2023evx} in a similar context).

Central to allow a low $\Lambda_{[23]}$ scale in Model B is the fact that the accidental $U(2)^5$ symmetry, 
which is present in both models below the $\Lambda_{12}$ scale, in  Model B
is broken in a minimal way: it is broken only by the $\psi^3_L \to \psi^{\rm light}_L$  mixing and by the 2$\times$2 light sub-blocks of the Yukawa coupling matrices. As discussed in~\cite{Barbieri:2011ci}, this minimal breaking implies FCNC amplitudes in the quark sector with the same CKM suppression as in the SM. 
The separation of the mechanism of flavour mixing (confined only to the left-handed sector) from that of the light-quark Yukawa couplings, 
is essential to achieve this goal. This does not happens in Model A, which is apparently very similar to Model B, is equally successful in describing the SM spectrum, and has a more minimal field content (same number of fermionic degrees of freedom and more economical Higgs sector).
The non-minimal $U(2)^5$--breaking terms present in this model at the $\Lambda_{[23]}$  scale  push the latter above 10~TeV.

A further interesting aspect which emerges by the explicit analysis of the two models is the sensitivity of flavour-changing processes
involving only light fields, 
such as $K$--$\bar K$ mixing and  $\mu \to e \gamma$,  to physics related to the $\Lambda_{12}$ scale, which is necessarily above 
100~TeV. While this is naturally expected by EFT considerations, it is interesting to note that even in the case of the Model~B rare processes involving only light fields provide stringent constraints on the flavour symmetry breaking occurring at the $\Lambda_{12}$ scale.
More generally, the analysis reported in Section~\ref{BSNP} illustrates the potential of the precision program 
in flavour physics of the next $10\div 15$ years, including the searches for lepton flavour violation in charged leptons,
where several observables/limits are expected to improve in precision by 
one order of magnitude or more.
 
In conclusion, it is worth noting that the gauge group in Eq.~(\ref{eq:G}) is a subset of the more ambitious unified groups presented in  Ref.~\cite{Davighi:2023iks}, which could provide a natural embedding of our two models. Indeed, what we have presented does not claim to be the definitive UV completion of the SM, nor the ultimate theory of flavour.  However, it does represents a sufficiently general set-up for investigating in details 
the non-trivial aspects associated with an explicit gauge theory capable of describing all the fermion masses, including those  of the light fermions, without small couplings. It also represents an efficient tool for analysing the interplay of different observables in probing this class of models, with only some  of the highlights discussed in this paper.   Needless to say, many other issues remain to be investigated, like, e.g., the maximal UV scale consistent with perturbativity and
electroweak vacuum stability.

\section*{Acknowledgments}
We would like to thank Joe Davighi for useful discussions.
This project has received funding from the European Research Council~(ERC) under the European Union's Horizon~2020 research and innovation programme under grant agreement 833280~(FLAY), and by the Swiss National Science Foundation~(SNF) under contract~200020\_204428. 

\section*{Appendix: Problems with alternative $U(1)$ groups}
\label{app:A}

The choice of the optimal gauge group to construct a multi-scale flavour theory  is a relevant open question. Here we insist that $SU(3)\times SU(2)$ act universally on the three families of chiral fermions and that only the third generation be directly coupled to the Higgs field(s).  This leaves open the choice of the $U(1)$ factors of the gauge group,  as well as the additional particle content.

Let us consider first the  more economical $U(1)_Y^{[3]} \times U(1)_Y^{[2]} \times U(1)_Y^{[1]}$ possibility~\cite{FernandezNavarro:2023rhv} and, in particular, the charged lepton sector.  Insisting on A-like models, where the additional fields are vector-like $SU(2)$-singlet fermions, it is easy to see how the Higgs field and the needed three electron-type vector-like fermions have to transform under $U(1)_Y^{[3]} \times U(1)_Y^{[2]} \times U(1)_Y^{[1]}$, as shown
in Table \ref{U1cube}.  However, a problem arises from the fact that the needed  mass mixing $\bar{E}_3e_1$ transforms  under the $U(1)$ factors as $(0, 1/2, -1/2)$, i.e. in the same way as $\bar{E}_2 E_1$. This does not allow 
the distinction between the first and the second generation, which have to get their masses at the higher and at the intermediate scale respectively.

This problem can be circumvented by leaving out $E_1$ and $E_2$ and by adding two  $SU(2)$-doublets $H_{u,d}^{[12]}$ with $Y^{[2]}=-1/2$ and coupled to $H_{u,d}$ as in Model B. The elimination of $E_{1,2}$ gives rise to accidental flavour conservation in the charged lepton sector, which is acceptable since lepton flavour conservation can be broken in the neutrino sector, but has to be avoided in the quark sector. However, as readily seen by a general spurion analysis, quark-flavour breaking (QFB) cannot be implemented with the scalar content of Table \ref{tab:SSB} augmented with $H_{u,d}^{[12]}$ and with hypercharges $Y^{(i)}, i=1,2,3$, suitably redefined. To implement QFB one needs to introduce at least two {\it ad hoc} new scalars $\eta, \hat{\eta}$ with the quantum numbers shown in Table \ref{tab:scalarsC} and the breaking pattern
\begin{equation}
U(1)_Y^{[3]} \times U(1)_Y^{[2]} \times U(1)_Y^{[1]} 
\stackrel{\langle\sigma\rangle}{\longrightarrow}
U(1)_Y^{[3]} \times U(1)_{Y}^{[12]} 
\stackrel{\langle\phi,\chi,\eta, \hat{\eta}\rangle}{\longrightarrow}
 U(1)_Y.
 \label{Y3-break}
\end{equation}
In this way one gets
\begin{equation}
Y^u =\left(\begin{array}{ccc}
	O(\epsilon_\sigma\epsilon_\phi)&  \approx 0 &  O(\epsilon_{\hat{\eta}})	\\
	\approx 0&  O(\epsilon_\phi)&  O(\epsilon_\chi) 	\\
	\approx 0&  O(\epsilon_\phi\epsilon_\chi)& O(1) 
\end{array}\right)\,, \qquad 
Y^d =\left(\begin{array}{ccc}
	O(\epsilon_\sigma\epsilon_\phi)& O(\epsilon_\eta)  &  O(\epsilon_{\hat{\eta}})	\\
	\approx 0&  O(\epsilon_\phi)&  O(\epsilon_\chi) 	\\
	\approx 0&  O(\epsilon_\phi\epsilon_\chi)& O(1) 
\end{array}\right)\,,
\label{eq:Y_C}
\end{equation}
where, on top of (\ref{eq:epsilon}), one has to introduce as well
\begin{equation}
\epsilon_{\eta}=\frac{ \langle \eta \rangle}{ \Lambda_{[23]} } ,\quad\quad \epsilon_{\hat{\eta}} =\frac{ \langle \hat{\eta }\rangle}{ \Lambda_{[23]} }
\end{equation}
A VL fermion content that would allow the explicit definition of the coefficients in Eq.~(\ref{eq:Y_C}) is given in Table \ref{tab:fermionsC}.

We find Eq.~(\ref{eq:Y_C}) less compelling than Eqs.~(\ref{eq:YuA}) and (\ref{eq:YuB}).
Note that, after inclusion of all $d=4$ gauge invariant terms in the potential for the scalar fields 
in Table~\ref{tab:scalarsC}, the last stage of breaking (\ref{Y3-break}) gives rise to a pseudo-Goldstone boson 
which gets a mass of $O(v)$ at electroweak symmetry breaking.

 \begin{table}[t]
 $$\begin{array}{c|c|c|c}
 & U(1)_Y^{[3]} & U(1)_Y^{[2]}& U(1)_Y^{[1]}\\ \hline
H_{u,d}& -1/2& 0& 0  \\ \hline\hline
E_2 & -1/2& -1/2& 0  \\ \hline
E_1 & -1/2&0 & -1/2 \\ \hline \hline
E_3 & 0& -1/2& - 1/2  \\ \hline
\end{array}$$
\caption{\small  Transformation properties of the Higgs fields and of the vector-like charged leptons in a $U(1)_Y^{[3]} \times U(1)_Y^{[2]} \times U(1)_Y^{[1]}$ model.}
\label{U1cube}
 $$\begin{array}{c|c|c|c|c}
 & U(1)_Y^{[3]} & U(1)_Y^{[2]}  & U(1)_Y^{[1]} &  SU(3)\times SU(2)\\ \hline
 H_{u,d} &  -1/2 & 0 & 0&(\bold{1},\bold{2}) \\ \hline \hline
 H^{[12]}_{u,d} & 0 &  -1/2 & 0 &(\bold{1},\bold{2}) \\ \hline 
\phi  & 1/2 & -1/2 & 0 &(\bold{1},\bold{1}) \\ \hline
\chi  & 1/6 & -1/6 & 0 &(\bold{1},\bold{1}) \\ \hline
\eta  & 1/2 & -1/3 & -1/6  &(\bold{1},\bold{1}) \\ \hline
\hat{\eta}  & 1/6 & 0 & -1/6  &(\bold{1},\bold{1}) \\ \hline
 \hline
\sigma  & 0 & -1/2 & 1/2  &(\bold{1},\bold{1}) \\ \hline
\end{array}$$
\caption{\small  Scalar content of a  $U(1)_Y^{[3]} \times U(1)_Y^{[2]} \times U(1)_Y^{[1]}$ model.}
\label{tab:scalarsC}
 $$\begin{array}{c|c|c|c|c}
 & U(1)_Y^{[3]} & U(1)_Y^{[2]}  & U(1)_Y^{[1]} &  SU(3)\times SU(2)\\ \hline
U_1 &  1/2 & 0 & 1/6&(\bold{3},\bold{1}) \\ \hline 
 U_2& 1/2&  1/6& 0 &(\bold{3},\bold{1}) \\ \hline 
D_1 & -1/2 & 0 & 1/6 &(\bold{3},\bold{1}) \\ \hline
D_2& -1/2 & 1/6 & 0 &(\bold{3},\bold{1}) \\ \hline \hline
U_3 &0 &  1/2& 1/6  &(\bold{3},\bold{1}) \\ \hline
D_3  & 0& -1/2& 1/6  &(\bold{3},\bold{1}) \\ \hline
E_3  &0& -1/2& -1/2 &(\bold{1},\bold{1}) \\ \hline
\end{array}$$
\caption{\small  VL fermion content of a  $U(1)_Y^{[3]} \times U(1)_Y^{[2]} \times U(1)_Y^{[1]}$ model.}
\label{tab:fermionsC}
\end{table}

Going to four $U(1)$ factors of the gauge group, as in Model A or B, there are three more choices that are anomaly free. 
The choices $U(1)_{T_{3R}} \times U(1)_{B-L}^{[3]} \times U(1)_{B-L}^{[2]}  \times U(1)_{B-L}^{[1]} $ or $U(1)_{B-L} \times U(1)_{T_{3R}}^{[3]}  \times U(1)_{T_{3R}}^{[2]} \times U(1)_{T_{3R}}^{[1]}$ do not work since, in either cases, it is not possible to fix the $U(1)$ charges of the Higgs bosons in such a way that they be directly coupled to the third generation only. 

Finally, in the case of $U(1)_Y^{[3]} \times U(1)_{T_{3R}}^{[12]} \times U(1)_{B-L}^{[2]} \times U(1)_{B-L}^{[1]}$, once the $U(1)$ charges of the vector-like $U_\alpha$ fermions are chosen in such a way  that every $q_i, i=1,2,3$ be coupled to the Higgs boson, as in Model A, one finds that the mixing terms $\bar{U}_2u_2$  and $\bar{U}_1u_1$ transform in the same way under the full gauge group, thus not making possible to distinguish between the second and the first fermion generation.

\bibliographystyle{JHEP}
\bibliography{references}
\end{document}